\documentclass[apj]{emulateapj}

\newcommand {\Ha}     {H$\alpha$}
\newcommand {\HH}     {H$_2$}        
\newcommand {\HI}     {\ion{H}{1}}   
\newcommand {\HII}    {\ion{H}{2}}   
\newcommand {\OVI}    {\ion{O}{6}}   
\newcommand {\OI}     {\ion{O}{1}}
\newcommand {\CIV}    {\ion{C}{4}}   
\newcommand {\NV}     {\ion{N}{5}}
\newcommand {\SiIV}   {\ion{Si}{4}}
\newcommand {\SiII}   {\ion{Si}{2}}

\newcommand {\FeII}   {\ion{Fe}{2}}

\newcommand {\lam}    {$\lambda$}
\newcommand {\kms}    {km~s$^{-1}$}
\newcommand {\NOVI}   {$N_{\rm OVI}$}

\newcommand\sk[2]{Sk\,{$-#1{^\circ}#2$}}
\newcommand {\tnma}   {\tablenotemark{a}}
\newcommand {\tnmb}   {\tablenotemark{b}}
\newcommand {\tnmc}   {\tablenotemark{c}}
\newcommand {\FUSE}   {{\it FUSE}}
\newcommand {\meanb}  {\langle b \rangle}
\newcommand {\meanv}  {\langle v \rangle}

\begin{document}

\title{Possible Detection of O\,VI from the Large Magellanic Cloud Superbubble N70}
\author{Charles W. Danforth}\affil{Center for Astrophysics \& Space Astronomy, Department of Astrophysical and Planetary Sciences, University of Colorado, 389-UCB, Boulder, CO 80309; danforth@casa.colorado.edu}
\author{William P. Blair}\affil{Department of Physics \& Astronomy, The Johns Hopkins University, 3400 N. Charles Street, Baltimore, MD 21218; wpb@pha.jhu.edu}

\begin{abstract}
We present {\it Far Ultraviolet Spectroscopic Explorer} (\FUSE) observations toward four stars in the Large Magellanic Cloud (LMC) superbubble N70 and compare these spectra to those of four comparison targets located in nearby field and diffuse regions.  The N70 sight lines show \OVI\ \lam1032 absorption that is consistently stronger than the comparison sight lines by $\sim60$\%.  We attribute the excess column density ($N_{\rm OVI}=10^{14.03}$~cm$^{-2}$) to hot gas within N70, potentially the first detection of \OVI\ associated with a superbubble.  In a survey of 12 LMC sight lines, Howk et al. (2002a) concluded that there was no correlation between ISM morphology and \NOVI.  We present a reanalysis of their measurements combined with our own and find a clear difference between the superbubble and field samples.  The five superbubbles probed to date with \FUSE\ show a consistently higher mean \NOVI\ than the 12 non-superbubble sight lines, though both samples show equivalent scatter from halo variability.  Possible ionization mechanisms for N70 are discussed, and we conclude that the observed \OVI\ could be the product of thermal conduction at the interface between the hot, X-ray emitting gas inside the superbubble and the cooler, photoionized material making up the shell seen prominently in \Ha.  We calculate the total hydrogen density $n_{\rm H}$ implied by our \OVI\ measurements and find a value consistent with expectations.  Finally, we discuss emission-line observations of \OVI\ from N70.
\end{abstract}


\section{Introduction}

Superbubbles (SBs) are some of the most dramatic examples of the interaction between hot stars and the surrounding interstellar medium (ISM).  They usually appear as limb-brightened rings $\sim100$ pc in diameter in optical emission lines, often surrounding stellar OB associations.  The combined stellar winds of the most massive association members excavate a bubble in the surrounding ISM and form a characteristic shell-like structure.  These bubbles are further heated and energized as the most massive association members reach the end of their lives and explode as supernovae (SN).  The coronal gas in the centers of SBs often appears bright in soft X-rays while the outer shell is bright in optical emission lines.  However, little is known observationally about the interface between the hot interior and the cool shell.  Measuring the hot gas content of SBs will help constrain the hot gas production in the ISM, disk-halo feedback mechanisms, as well as refine the models of the life-cycle of SBs. 

The advent of the Far Ultraviolet Spectroscopic Explorer (\FUSE) satellite has made it possible to observe the \OVI\ resonance doublet ($\lambda\lambda1031.926,1037.617$), which is a sensitive probe of $T=10^5-10^6$~K gas.  Since O$^{5+}$ has an ionization potential of 114 eV, photoionized \OVI\ is negligible in most ISM contexts.  Moderate-velocity shocks, however, can produce \OVI\ \citep[$v_{\rm shock}\ga150$ \kms][]{HRH}.  Gas in the $\sim10^5$~K phase can also be formed at the interface between hot, coronal gas ($T>10^6$ K) and cooler material through thermal conduction \citep{CowieMcKee77,Weaver77} and turbulent mixing layers \citep{BegelmanFabian90,Slavin93}.   

The Magellanic Clouds contain a number of classic examples of SBs.  However, nearly every sight line toward the Magellanic Clouds shows \OVI\ absorption at both Galactic and Magellanic velocities \citep{Danforth02,Hoopes02,Howk02a}.  For instance, \citet{Howk02a} measured the Large Magellanic Cloud (LMC) \OVI\ absorption in the spectra of twelve carefully-chosen stars and found that \NOVI\ has very little correlation with warm-disk ISM morphology as traced by \Ha\ or soft X-ray brightness.  Secondly, there are variations in \NOVI\ column density by a factor of four or more between sight lines and these variations occur over relatively small angular scales.  The \OVI\ absorption profiles show little relationship to tracers of cooler ions and are usually centered at approximately $-30$ \kms\ in relation to warm-disk probes such as \FeII.  Furthermore, the \OVI\ profiles are all much broader than can be accounted for by simple thermal broadening, implying that they sample multiple hot gas systems or turbulent absorbers along each line of sight.  These results are generally consistent with the idea that the observed \OVI\ at LMC velocities is due primarily to a halo of hot gas surrounding the LMC and expanding toward us up to several 10s of \kms.

\begin{figure*} 
  \epsscale{.7}\plotone{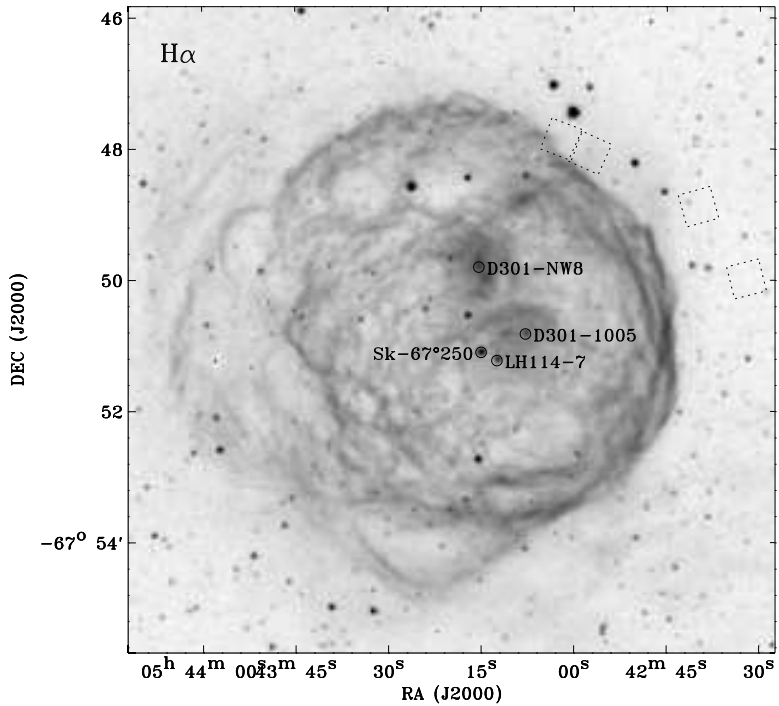} 
  \caption{N70 is a textbook example of an isolated superbubble, as
  seen in this H$\alpha$+[\ion{N}{2}] image.  Four stars observed with
  the \FUSE\ MDRS aperture lie clustered in the center of a
  limb-brightened ring $\sim$105 pc in diameter.  Boxes in the upper
  right show the locations of the corresponding serendipitous
  30\arcsec$\times$30\arcsec LWRS aperture positions during the stellar
  observations. (See text.)}
\end{figure*}

\begin{deluxetable*}{llrclclcc} 
  \tabletypesize{\footnotesize}\tablecolumns{9}\tablewidth{0pt} 
  \tablecaption{Log of \FUSE\ N70 and Comparison Observations}
  \tablehead{\colhead{Target}&\colhead{RA ~(J2000)~ Dec}&\colhead{N70}&
              \colhead{Spectral}&\colhead{V}&\colhead{$F_{1050}$\tnma}&
               \colhead{FUSE}&\colhead{Exp}&\colhead{Notes}\\ 
             \colhead{}&\colhead{h~~m~~s~~~~~~~~$^\circ$~~~\arcmin~~~\arcsec}~~&
              \colhead{dist.}&\colhead{Type}&\colhead{mag}&\colhead{}&\colhead{ID}&
               \colhead{(ksec)}&\colhead{}}
  \startdata
   LH\,114$-$7       &05 43 12.85 ~-67 41 16.2&\nodata  & O2\,Iab  &13.66 &120  &D09811& 13.9  &N70\\
   Sk\,$-67^\circ250$&05 43 15.48 ~-67 51 09.6&\nodata  & O9\,III  &12.68 &150  &D09812& 12.5  &N70\\
   D\,301$-$1005     &05 43 08.30 ~-67 50 52.4&\nodata  & O9\,V    &14.11 & 50  &D09814&  3.1  &N70\\
   D\,301$-$NW8      &05 43 15.96 ~-67 49 51.0&\nodata  & O7\,V    &14.37 & 40  &D09815& 12.2  &N70\\
   &&&&&&&\\	                         			    
   BI\,272  &05 44 23.18 ~-67 14 29.3&37\arcmin& O7:\,II-III:\tnmb &13.28 &130  &P11729& 10.9  &DEM\,310\\
   Sk\,$-67^\circ266$&05 45 52.00 ~-67 14 25.0&39\arcmin& O8:\,Iab:&12.01 &170  &B09002&  4.9  &DEM\,310\\
   BI\,237           &05 36 14.68 ~-67 39 19.3&41\arcmin& O3\,V    &13.86 & 25  &E51140& 35.6  &diffuse H\,II\\
   CAL\,83           &05 43 34.20 ~-68 22 22.0&31\arcmin& CV/XRB   &16.20 &  1.5&B01501& 45.8\tnmc&field
  \enddata
  \tablenotetext{a}{Approximate continuum flux at $\lambda\approx1050$
  \AA\ in units of ($\rm 10^{-14}~ergs~cm^{-2}~s^{-1}~\AA^{-1}$).}
  \tablenotetext{b}{\citet{Massa03} note uncertainty in spectral type; wind lines suggest this star may be hotter and closer to the main sequence.}
  \tablenotetext{c}{LiF1 and LiF2 data were coadded for this sight
  line for an effective exposure time of $\sim90$ ksec.}
\end{deluxetable*}

In this paper, we use \FUSE\ observations of a carefully selected set of stars within and adjacent to the LMC SB N70 \citep{Henize56} to disentangle the LMC halo and SB hot gas signatures.  The LMC is a nearly face-on disk galaxy \citep[$i=35\pm6^\circ$,][]{vanderMarelCioni01} at a well-known distance \citep[$\sim51$ kpc,][]{Cioni00} and free from the line-of-sight confusion and relative distance uncertainties present when comparing most Galactic sight lines.  

\begin{figure*} 
  \epsscale{.7}\plotone{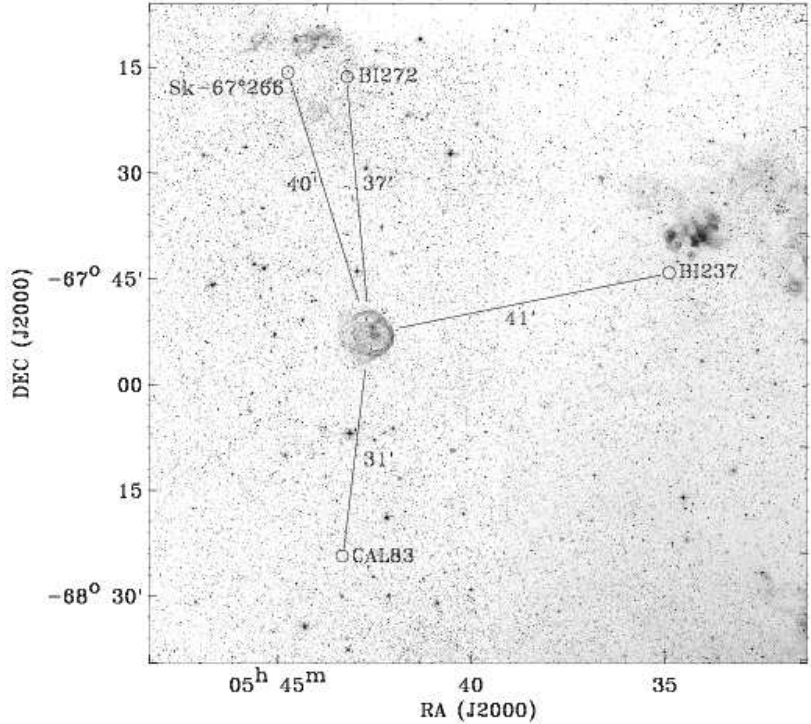} 
  \caption{The N70 sight lines are compared to four nearby sight lines
  toward hot stars.  The comparison stars are located in the field
  and diffuse H\,II emission on the eastern edge of the LMC.  Angular
  separations from N70 are overlaid on a POSS2/UKSTU red-band image.
  The rotational center of the LMC lies 2.6$^\circ$ west of N70.}
\end{figure*}

\section{Observations \& Data Analysis}

We chose N70 \citep[DEM\,301,][]{Henize56,DEMref} as our ideal candidate SB to probe for hot gas.  It is a beautiful ring nebula $\sim105$ pc in diameter centered 2.6$^\circ$ east of the dynamic center of the LMC (see Figure~1).  While the \HI\ and stellar components of the LMC are known to extend to large distances \citep{Kim99,vanderMarelCioni01}, the \Ha\ emission tends to be more concentrated \citep{SHASSAref}.   N70 is one of the most outlying bright emission-line objects,  located in an isolated region of the disk far from supergiant shells \citep{Meaburn80}, \HII\ regions, and bright diffuse \Ha\ emission.  The emission nebula is centered on an OB association (LH\,114) consisting of at least nine stars of type B0 or earlier \citep{Oey96a}.  An assumed Salpeter initial mass function predicts one additional $\rm\sim60\,M_\sun$ member which has presumably already exploded as a supernova \citep{Oey96b}.  The SB is consistent with modest shock heating, with a slightly enhanced [\ion{S}{2}]/\Ha\ ratio, weak X-ray emission, marginally non-thermal radio continuum, and a mildly supersonic expansion velocity \citep[$\sim 40$ \kms, see][and references therein]{ChuKennicutt88}.  N70 displays the same filamentary structure seen in many mature supernova remnants (SNRs) \citep{Skelton99}, but does not show any of the high-velocity emission knots commonly seen in those same SNRs \citep{Chu97}.  \citet{Oey00} model N70 as a density-bounded nebula with a SN shock and a photoionized post-shock gas arranged into the \Ha-bright filaments seen in the nebula.  From optical emission line ratios, they calculate that shock velocity must be $\sim70$ \kms, which is much too low to produce \OVI\ in standard radiative shock models. 

We obtained \FUSE\ observations toward four early-type stars projected within N70 to search for evidence of hot gas.  The \FUSE\ instrument covers a spectral range of $905-1187$ \AA\ (including the \OVI\ doublet) at a resolution of $\sim15$ \kms. Further details about the instrument and on-orbit performance can be found in \citet{Moos00} and \citet{Sahnow00}. 

The N70 targets are all of spectral type O9 or earlier and have well-characterized continua in the spectral region near the \OVI\ \lam1032 line.  The weaker \OVI\ \lam1038 line is blended with strong absorption from \OI, \HH, and \ion{C}{2}* ISM lines and is not usually usable.  The \FUSE\ targets are indicated in Figure~1 and the observational specifics are listed in Table~1. 

We gauge the \OVI\ contribution from the LMC halo in the direction of N70 by selecting a group of nearby comparison sight lines from archival \FUSE\ data.  Though \FUSE\ has observed over 100 early-type stars in the LMC \citep{Danforth02,DanforthThesis}, we require specific attributes for optimal comparison targets.  Proximity to N70 is the first requirement since we wish to probe a similar sight line through the LMC halo material;  we limit our search to observations within 60\arcmin\ of N70.  Furthermore, we wish to avoid contamination from possible local sources of hot gas and thus disqualify targets toward \HII\ regions, other SBs, supergiant shells, and SNRs.  Finally we need an unambiguous stellar continuum around 1032\AA\ and thus rule out stars of type B0 and later.  We found three O-type stars observed by \FUSE\ that match these criteria.  In addition, we included CAL\,83 (RX\,J0543.6-6822), a cataclysmic variable/X-ray binary target located in a the low-emission field south of N70.  This target is fainter, but the continuum is well-defined and acceptable for use as a comparison.  Figure~2 shows that our comparison stars are located on three sides of N70.  Thus, any large-scale trends in \OVI\ column density across the region should be averaged out to first order.  We also list the observational parameters for these comparison targets in Table~1.

Each of the eight data sets was reduced with {\sc CalFUSE v2.0.5} or higher, aligned, and coadded by exposure.  Our analysis is based on LiF1 data since this channel has the highest sensitivity.  The exception is CAL\,83, for which we aligned and coadded the LiF1a and LiF2b channels, resulting in an increased S/N for this faint source (at the cost of a slight decrease in spectral resolution).  \OVI\ absorption lines tend to be broad ($b\sim40$ \kms) compared to the instrumental profile ($b_{\rm inst}\sim10$ \kms) so this slight decrease in resolution is acceptable.  We converted observed heliocentric velocities to the local standard of rest (LSR) with a uniform correction of $v_{\rm LSR}=v_\sun-15.7$ \kms.  The wavelength scale of each sight line was then calibrated such that the strongest component of the narrow \SiII\ \lam1020.70 line was at $v_{\rm lsr}=0$ \kms, an adjustment of much less than a 20 \kms\ resolution element in all cases.  Data from the LiF1b segment were aligned with the LiF1a segment by cross-correlating \FeII\ \lam1144.94 absorption with \SiII\ \lam1020.70.  Again, this involved a shift of less than a resolution element in all cases.

\begin{figure*}  
  \epsscale{1}\plotone{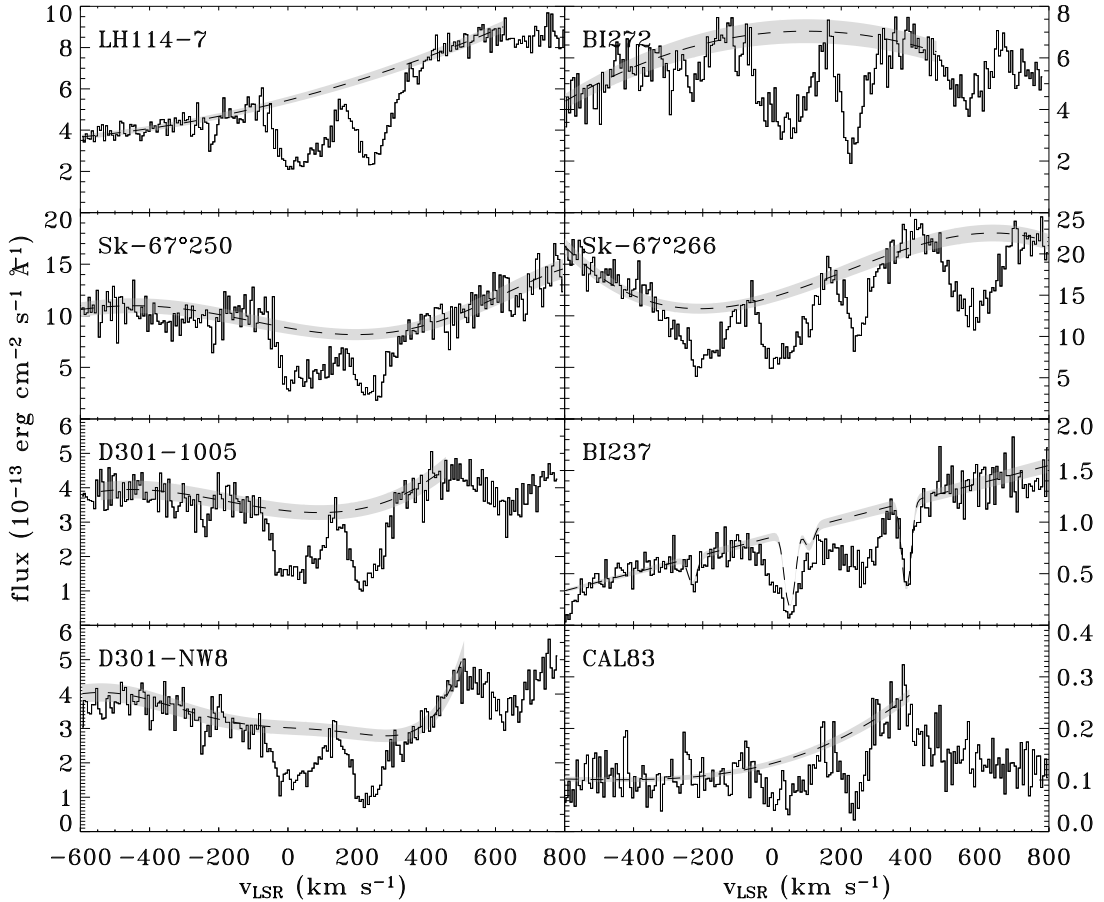} 
  \caption{O\,VI continuum fits for the N70 (left) and comparison
  (right) sight lines.  All data are from the LiF1a channel except
  CAL\,83 for which LiF1a and LiF2b have been aligned and coadded.
  The N70 sight lines were observed through the
  4\arcsec$\times$20\arcsec\ MDRS aperture while the comparison sight
  lines were all observed with the 30\arcsec$\times$30\arcsec\ LWRS
  aperture.  The absorption near zero velocity is attributed to
  Galactic O\,VI, and that near $+$240 \kms\ is due to O\,VI in the
  LMC.  Continuum fits (dashed) and $\pm1\sigma$ uncertainties
  (shaded) are low-order Legendre polynomial fits to absorption-free
  regions of the spectrum around 1032\AA.  The BI\,237 continuum
  includes modeled \HH\ absorption as described in the text.}
\end{figure*}

Absorption-free regions of stellar continuum within a few Angstroms of 1032 \AA\ were used to normalize the spectra with low-order Legendre polynomials.  The \OVI\ data and continuum fits are shown in Figure~3 for both the N70 and comparison sight lines.  The shaded regions denote the $\pm1\sigma$ range of continuum placements.  Note that the CAL83 continuum is generally well-defined except in the region of the stellar \OVI\ emission feature redward of the LMC absorption line.  We choose to fit a smooth curve to the blue side of the emission line and ignore continuum longward of this feature.

\begin{deluxetable}{cccc} 
  \tabletypesize{\footnotesize}\tablecolumns{4}\tablewidth{0pt} 
  \tablecaption{\HH\ Measurements for BI\,237}
  \tablehead{\colhead{J} & \colhead{$v$} & \colhead{$b$} & \colhead{log\,$N$} }
  \startdata
    J=3 & $+5\pm5$  & $15\pm5$ & $14.7\pm0.2$    \\
        & $281\pm1$ & $16\pm2$ & $15.24\pm0.10$ \\
    J=4 & $\sim-5$  & $\sim16$ & $<14.2$         \\
        & $281\pm1$ & $14\pm2$ & $14.90\pm0.12$
  \enddata
\end{deluxetable}

Several \HH\ lines can potentially contaminate the \OVI\ \lam1032 profile, specifically absorption at Galactic and LMC velocities in the (6-0)P3\lam1031.191 and (6-0)R4\lam1032.349 transitions.  Fortunately, absorption in $J\ge3$ is less common than for $J<3$ toward the LMC \citep{Tumlinson02}.  The only sight line with significant $J\ge3$ \HH\ absorption is BI\,237.  We measured apparent column densities, line widths, and velocities in nine different unblended, $J=3,4$ \HH\ lines in the BI\,237 data and averaged the results for each $J$-state and velocity (see Table~2).  A model \HH\ spectrum was produced based on these values and convolved with the \FUSE\ aperture function.  The resulting model was divided from the observed \OVI+\HH\ profile around 1032 \AA\ in the BI\,237 data to derive a corrected profile.  The \HH\ model is shown in the BI\,237 continuum fit in Figure~3.

\begin{deluxetable}{lcccl} 
  \tabletypesize{\footnotesize}
  \tablecolumns{5}
  \tablewidth{0pt} 
  \tablecaption{O\,VI \lam1032 Measurements}
  \tablehead{\colhead{Target}&
	\colhead{$v$ limits\tnma}&
	\colhead{$\meanv$\tnma}&
	\colhead{$\meanb$\tnma}&
	\colhead{log\,$N_a$} }
  \startdata
  \cutinhead{Galactic Absorption, AOD Method}
    LH\,114-7          & -55, 154 &$  44\pm 2$ &$ 74\pm2$ &$ 14.56\pm0.02$         \\
    Sk\,$-67^\circ250$ & -50, 156 &$  53\pm 3$ &$ 76\pm4$ &$ 14.60\pm0.04$         \\
    D\,301$-$1005      & -66, 153 &$  39\pm 1$ &$ 68\pm1$ &$ 14.49\pm0.04$         \\
    D\,301-NW8         & -53, 155 &$  39\pm 2$ &$ 63\pm4$ &$ 14.38^{+0.06}_{-0.07}$\\
    BI\,272            & -55, 174 &$  49\pm 0$ &$ 74\pm1$ &$ 14.47\pm0.03$         \\
    Sk\,$-67^\circ266$ & -58, 153 &$  33\pm 1$ &$ 66\pm2$ &$ 14.41\pm0.04$         \\
    BI\,237            & -73, 124 &$  31\pm 1$ &$ 63\pm1$ &$ 14.47^{+0.06}_{-0.08}$\\
    CAL\,83            & -72, 137 &$  21\pm 1$ &$ 56\pm5$ &$ 14.43^{+0.08}_{-0.09}$\\
 $\langle$N70$\rangle$ & \nodata  &$  44\pm 7$ &$ 70\pm6$ &$ 14.51\pm0.10\pm0.04$\tnmb\\
 $\langle$comp$\rangle$& \nodata  &$  34\pm12$ &$ 65\pm7$ &$ 14.45\pm0.03\pm0.06$\tnmb\\
  \cutinhead{LMC Absorption, AOD Method}
    LH\,114-7          & 154, 350 &$ 243\pm 2$ &$ 61\pm 2$ &$ 14.45\pm0.02$         \\
    Sk\,$-67^\circ250$ & 156, 340 &$ 237\pm 3$ &$ 59\pm 3$ &$ 14.57\pm0.05$         \\
    D\,301$-$1005      & 153, 323 &$ 241\pm 0$ &$ 54\pm 1$ &$ 14.45\pm0.04$         \\
    D\,301-NW8         & 155, 326 &$ 248\pm 1$ &$ 48\pm 4$ &$ 14.45^{+0.05}_{-0.06}$\\
    BI\,272            & 174, 328 &$ 245\pm 0$ &$ 45\pm 1$ &$ 14.30\pm0.03$         \\
    Sk\,$-67^\circ266$ & 153, 325 &$ 242\pm 1$ &$ 41\pm 6$ &$ 14.09^{+0.06}_{-0.07}$\\
    BI\,237            & 124, 336 &$ 232\pm 1$ &$ 70\pm 2$ &$ 14.39\pm0.03$         \\
    CAL\,83            & 137, 287 &$ 212\pm 0$ &$ 44\pm 2$ &$ 14.39^{+0.07}_{-0.08}$\\
 $\langle$N70$\rangle$ & \nodata  &$ 242\pm 5$ &$ 56\pm 6$ &$ 14.48\pm0.06\pm0.04$\tnmb\\
 $\langle$comp$\rangle$& \nodata  &$ 233\pm15$ &$ 50\pm13$ &$ 14.29\pm0.14\pm0.05$\tnmb\\
  \cutinhead{LMC Absorption, Profile Fit Method} 	     
    LH\,114-7          & \nodata  &$ 241\pm2 $ &$ 61\pm 4$ &$ 14.46\pm0.03$         \\
    Sk\,$-67^\circ250$ & \nodata  &$ 234\pm4 $ &$ 58\pm 6$ &$ 14.61^{+0.04}_{-0.05}$\\   
    D\,301$-$1005      & \nodata  &$ 241\pm2 $ &$ 52\pm 4$ &$ 14.46^{+0.04}_{-0.03}$\\  
    D\,301-NW8         & \nodata  &$ 247\pm3 $ &$ 49\pm 4$ &$ 14.47^{+0.06}_{-0.05}$\\   
    BI\,272            & \nodata  &$ 240\pm2 $ &$ 39\pm 3$ &$ 14.30\pm0.04$         \\
    Sk\,$-67^\circ266$ & \nodata  &$ 239\pm3 $ &$ 38\pm 4$ &$ 14.09\pm0.06$         \\   
    BI\,237            & \nodata  &$ 240\pm6 $ &$ 72\pm 9$ &$ 14.40\pm0.05$         \\   
    CAL\,83            & \nodata  &$ 222\pm7 $ &$ 34\pm 9$ &$ 14.37^{+0.12}_{-0.13}$\\   
 $\langle$N70$\rangle$ & \nodata  &$ 241\pm5 $ &$ 55\pm 5$ &$ 14.50\pm0.07\pm0.04$\tnmb\\
 $\langle$comp$\rangle$& \nodata  &$ 235\pm9 $ &$ 46\pm18$ &$ 14.29\pm0.14\pm0.07$\tnmb\\
  \enddata
  \tablenotetext{a}{Units are \kms.}
  \tablenotetext{b}{First quoted uncertainty is $1\sigma$ deviation of mean
  values.  Second uncertainty is mean of measured errors.}
\end{deluxetable}

The Galactic ($v\la150$ \kms) and LMC ($v\ga150$ \kms) absorption components are distinct and well-resolved by \FUSE\ for all sight lines even though there may be multiple absorption subcomponents hidden in each system.  We measure the apparent optical depth (AOD) column density \citep{SavageSembach91} and profile-weighted mean velocity and line width \citep{SembachSavage92} for both Galactic and LMC absorption (Table~3).  Saturation is a concern with the AOD method and the apparent column density $N_a$ is usually taken as a lower limit to the true column density $N$.  In this case, we are reasonably certain that the \OVI\ absorption is not saturated;  the profiles are broad and do not approach zero.  The thermal line width $b_{\rm therm}=17$ \kms\ for \OVI\ at its peak collisional ionization equilibrium temperature of $T=10^{5.45}$~K, and this width is comparable to the \FUSE\ resolution element.  

Uncertainties in our measurements arise from continuum placement error and velocity integration limit uncertainty.  To properly address this, we vary the continuum placement by $\pm1\sigma$ and adjust the central velocity limit by $\pm10$ \kms.  These errors are added in quadrature to give confidence limits on our AOD measurements.  Finally, we calculate ensemble averages and standard deviations for all quantities (Table~3).

As a check on our AOD measurements, we fit the normalized spectra with a series of Voigt profile components convolved with the \FUSE\ instrumental profile with free parameters $v$, $b$, and $N$.  Component centroids were picked interactively and allowed to vary $\pm20$ \kms.  Doppler width $b$ is allowed to vary between $10\leq b\leq100$ \kms.  Uncertainties are the quadratic sum of fit uncertainties and high-low continuum placement as noted above.  The Galactic \OVI\ absorption is best fit with a pair of Voigt components at $v_{\rm LSR}\sim0,65$ \kms\ for most sight lines.  However, these two components are too closely blended for strong constraints on their individual fit parameters.  The sum of the two components is generally consistent with the AOD \NOVI\ measurements, but the resulting errors from the Voigt fit are very large and we do not consider these further.  The LMC \OVI\ absorption is generally well-fit with a single Voigt component at $v_{\rm LSR}\sim235$ \kms.  These fits (Table~3) are entirely consistent with the AOD measurements in $v$, $b$, and \NOVI\ with comparable uncertainties on all quantities and provide a valuable consistency check in our methodology.

\section{Discussion}

The measurements in Table~3 show an excess of \OVI\ at LMC velocities in the N70 sight lines compared with the field sight lines.  The mean, integrated apparent \OVI\ column density for the N70 sight lines is $\langle{\rm log}\,N_{\rm N70}\rangle=14.48\pm0.06$ while the comparison targets show $\langle{\rm log}\,N_{\rm comp}\rangle=14.29\pm0.14$.  The component fit measurements are nearly identical; $\langle{\rm log}\,N_{\rm N70}\rangle=14.50\pm0.07$ versus $\langle{\rm log}\,N_{\rm comp}\rangle=14.29\pm0.14$.  In either case, the N70 targets show $\sim60\%$ more \OVI\ column density ($N\rm_{OVI}=10^{14.03}~cm^{-2}$) at LMC velocities than the comparison targets, a difference larger than the dispersion in either sample.  The same is not true of the Galactic absorption; the mean integrated Galactic \OVI\ columns are $\langle{\rm log}\,N\rangle=14.51\pm0.10$ and $\langle{\rm log}\,N\rangle=14.45\pm0.03$ for the N70 and field sight lines, respectively.  While the N70 sight lines show a slight excess in comparison to the field sight lines, it is only by a factor of $15\%$ and is within the errors.

Kinematically, we see that the N70 \OVI\ profiles and comparison profiles have consistent LMC absorption centroids ($\meanv_{\rm N70}=242\pm5$ \kms\ and $\meanv_{\rm comp}=233\pm15$ \kms) and line widths ($\meanb_{\rm N70}=56\pm6$ \kms and $\meanb_{\rm comp}=50\pm13$ \kms).  The N70 absorption is marginally broader and at higher velocity than the comparison absorption, but this is within the error bars in both cases.    

\begin{figure} 
  \epsscale{1.2}\plotone{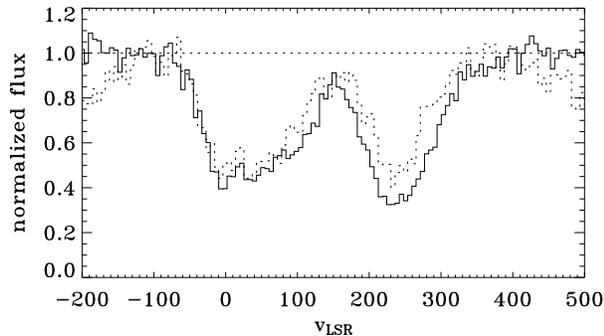} 
  \caption{Mean composite O\,VI profiles for N70 (solid) and the
  comparison sight lines (dotted).  The two profiles show remarkably
  similar structure in the Galactic component, but the LMC absorption
  in N70 noticeably stronger and broader.}
\end{figure}

To illustrate the differences between SB and comparison profiles graphically, we combined the normalized \OVI\ data for the N70 and comparison sight lines pixel-by-pixel with an error-weighted average to create composite profiles.  Combining the N70 sight lines makes sense as they are all within a few arcminutes of each other and probe nearly identical lines of sight.  The comparison sight lines are separated by more than a degree on the sky in some cases and probe potentially different material, particularly at the distance of the LMC.  Still, the scatter in comparison sight line centroid velocities is below the \FUSE\ resolution (comparable to or smaller than $\sigma(v)$ in the N70 sight lines) and the data were deemed homogenous enough to combine into a composite profile.  From Figure~4, we see that the Galactic profiles are very similar in shape and that the N70 profile shows only slightly greater depth.  The LMC components are centered at nearly the same velocity, but show decidedly different average profiles.  We note that these composite profiles are used for illustrative purposes only and that all quoted measurements are averages of individual sight line data, not the combined data. 

Is the 60\% excess \OVI\ column density observed in the N70 profiles compared to nearby stars due to excess \OVI\ within N70 or a fluctuation between LMC halo sight lines?  \citet{Howk02a} measured \OVI\ absorption in 12 selected sight lines toward the LMC.  These sight lines probed many different ISM morphologies, but they found no correlation between \NOVI\ and ISM morphology as traced by either local \Ha\ or X-ray brightness.  Similarly, they found no correlation between \NOVI\ and the spectral type of the star observed by \FUSE\ (usually the most massive star in the local OB-association).  They found that \NOVI\ varies by a factor of $\sim4$ between sight lines with a mean of $\langle{\rm log}N_{\rm OVI}\rangle=14.37^{+0.14}_{-0.21}$ ($1\sigma$ deviations).  Despite the excess \OVI\ seen in our N70 observations, the mean column densities for both N70 and the comparison sample fall within this range.

\begin{deluxetable}{lcllc} 
  \tabletypesize{\footnotesize}\tablecolumns{5}\tablewidth{0pt} 
  \tablecaption{LMC \NOVI\ Summary}
  \tablehead{\colhead{Sample}             &
             \colhead{${\cal N}$}           &
             \colhead{$\langle {\rm log} N_{\rm OVI}\rangle$}&
	     \colhead{$\langle N_{\rm OVI}\rangle/10^{14}$}&
             \colhead{Ref.\tnma}}
  \startdata
    Howk et al. sample          & 12 & $14.37^{+0.14}_{-0.21}$ & $2.34\pm0.89$ &  1   \\
    Howk et al. sample\tnmb     & 11 & $14.40^{+0.12}_{-0.16}$ & $2.49\pm0.78$ &  1   \\
    this work sample            &  8 & $14.39\pm0.14$          & $2.54\pm0.72$ &  2   \\
    total LMC sample            & 20 & $14.35\pm0.17$          & $2.42\pm0.81$ &  1,2 \\
    total LMC sample\tnmb       & 19 & $14.38\pm0.13$          & $2.51\pm0.73$ &  1,2 \\
&&&\\				        		                       
    Howk et al. SBs             &  4 & $14.49\pm0.12$          & $3.10\pm0.84$ &  1   \\
    N70 sight lines             &  4 & $14.48\pm0.06$          & $3.04\pm0.45$ &  2   \\
    total SB sample           &8\tnmc& $14.48\pm0.09$          & $3.07\pm0.62$ &  1,2 \\
&&&\\				        		                       
    Howk et al. non-SBs         &  8 & $14.29\pm0.18$          & $1.96\pm0.66$ &  1   \\
    Howk et al. non-SBs\tnmb    &  7 & $14.32\pm0.09$          & $2.13\pm0.49$ &  1   \\
    comparison sight lines      &  4 & $14.29\pm0.14$          & $2.03\pm0.58$ &  2   \\
    total non-SB sample         & 12 & $14.28\pm0.16$          & $1.99\pm0.61$ &  1,2 \\
    total non-SB sample\tnmb    & 11 & $14.31\pm0.11$          & $2.10\pm0.50$ &  1,2 \\
  \enddata
  \tablenotetext{a}{Sources: 1 - \citet{Howk02a}, 2 - this work}
  \tablenotetext{b}{Excluding Sk\,$-67^\circ05$ sight line, see \citet{Howk02a}}
  \tablenotetext{c}{Sight lines probe five SBs: N70, N144, N204, N206, and N154}
\end{deluxetable}

We have re-examined the \citet{Howk02a} sight lines, however, and reach a somewhat different conclusion.  Four of the stars used by Howk et al. are within SBs: \sk{68}{80} within superbubble N144, \sk{70}{91} (N204), \sk{71}{45} (N206), and \sk{69}{191} (N154).  These sight lines show $\langle{\rm log}\,N\rm_{OVI}(SB)\rangle=14.49\pm0.12$ while the other eight, non-SB sight lines show $\langle{\rm log}\,N\rm_{OVI}(non-SB)\rangle=14.29\pm0.18$ (see Table~4).  These column densities and the relative $\sim60\%$ difference between them are remarkably consistent with our measurements of the four N70 and four comparison sight lines, respectively.  This is strong corroborating evidence for a SB origin of the excess \NOVI.  Furthermore, this assessment raises hope that other SBs can be studied systematically for assessing \OVI\ absorption.

If this interpretation is correct, it has important implications for the hot gas in galactic halos.  First, the Howk et al. mean halo \NOVI\ is ``contaminated'' by contributions from four SBs.  Removing the SB sight lines (and including our four comparison measurements) results in a lower value for the LMC halo component (Table~4).  Surprisingly, excluding SB sight lines from the LMC halo survey does not appreciably reduce the degree of variability of halo \OVI.  The entire sample of 20 LMC sight lines (this work plus the Howk et al. sample) shows $\langle{\rm log}N_{\rm OVI}\rangle=14.35\pm0.17$ while the twelve non-SB sight lines show $\langle{\rm log}N_{\rm OVI}\rangle=14.28\pm0.16$.  Howk et al. note that one target (Sk\,$-67^\circ05$) has a questionable continuum placement and low wind velocity.  We note that this target shows by far the lowest \NOVI\ in the combined sample and is located on the extreme western edge of the LMC.  If this point is removed, the means become $\langle{\rm log}N_{\rm OVI}\rangle=14.38\pm0.13$ for the full 19-target sample and $\langle{\rm log}N_{\rm OVI}\rangle=14.31\pm0.11$ for the 11 field sight lines.  In both scenarios (with and without Sk\,$-67^\circ05$), the mean decreases significantly, but the dispersion remains essentially unchanged.  

The SB sample shows a systematically higher column density with $\langle{\rm log}N_{\rm OVI}\rangle=14.48\pm0.09$ for eight targets in five SBs.  While this is less variation in log\,\NOVI, it actually represents a similar absolute variation in linear \NOVI\ (see Table~4).  If the systematically higher \NOVI\ observed toward SB targets represents a the sum of halo \OVI\ plus a constant SB contribution, the halo contribution shows no less absolute variability than the non-SB targets.  Thus the sample of five SBs probed to-date in the LMC (eight sight lines) shows a significantly higher mean \NOVI\ than do the twelve non-SB sight lines.  The absolute scatter within the samples, however, is equivalent for SB and non-SBs.  The database of stellar LMC \FUSE\ observations of the earliest-type stars has grown substantially and a detailed analysis of additional sight lines would undoubtedly improve the statistics on SB/non-SB \OVI.

Howk et al. (2002a) may have been too conservative in their conclusion regarding the lack of correlation between \NOVI\ and ISM morphology.  They probably saw no correlation between \NOVI\ and \Ha\ brightness because they trace different processes;  \Ha\ brightness is largely a function of photoionization while \OVI\ is a product of collisional ionization and/or interfaces.  Both \HII\ regions and SBs are bright in \Ha\ emission, but only the latter have the conditions required for \OVI\ production.  The lack of correlation between stellar spectral type and \NOVI\ is also peripheral as SB dynamics and \OVI\ production are often driven by the evolved members of a cluster which have since exploded as SN, not the current association members observed with \FUSE. 

We measure a perpendicular column density in the Galactic halo component $\langle {\rm log} N_{\rm OVI}~{\rm sin}|b|\rangle=14.25\pm0.10$ for the N70 sight lines and $14.19\pm0.03$ for the comparison stars ($b\approx-33^\circ$).  Both values are consistent with larger studies of Galactic halo \OVI\ in extragalactic sight lines \citep{Howk02b,Savage03}, but a factor of three higher than that found in the first 10 kpc of the Galactic halo \citep{Zsargo03}.

\begin{figure} 
  \epsscale{1.2}\plotone{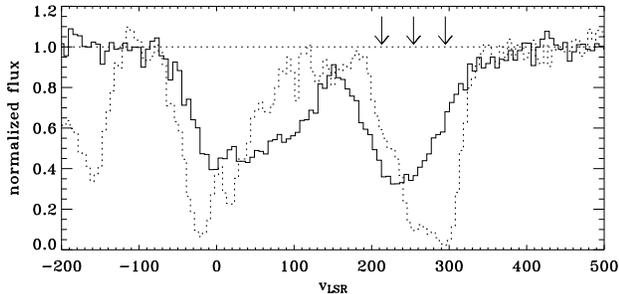} 
  \caption{Comparision of composite O\,VI $\lambda1032$ (solid) and
  Fe\,II $\lambda1145$ (dotted) profiles for the four N70 sight lines.
  The Fe\,II profile shows three distinct LMC components in all four
  sight lines at $v_{\rm}=213$, 254, and 295 \kms\ (arrows).}
\end{figure}

Absorption in the LMC disk is traced by low-ionization species such as \FeII, and the strongest \FeII\ component often indicates the systemic velocity at a given location in the LMC disk \citep{Danforth02,DanforthThesis}.  We normalized the strong \FeII\ $\lambda1145$ absorption for all four N70 sight lines and created a composite profile in the same manner as our composite \OVI\ profile described above.  Figure~5 shows \FeII\ in comparison to \OVI.  All four \FeII\ profiles show three distinct absorption components of increasing strength at $v_{\rm LSR}\approx213$, 254, and 295 \kms.  We take the strongest of these components ($v_{\rm LSR}\approx295$ \kms) to be warm disk gas at the N70 systemic velocity.  \citet{BEM80} measured optical [\ion{O}{2}] emission at various locations in N70 and see a strong emission peak at the same velocity at both eastern and western edges of the nebula.  We interpret the middle component ($v_{\rm LSR}=254$ \kms) as warm, photoionized gas in the approaching side of the SB, consistent with the quoted expansion velocity of $\sim40$ \kms.  Receding emission, presumably from the far side of N70, is seen by \citet{BEM80} at $v\sim311$ \kms\ toward the center of the SB.  We do not see any \FeII\ absorption near this velocity nor do we see \OVI\ absorption above the systemic velocity which supports our claim that the \FUSE\ target stars are located within the SB, not behind it.  

The weak $v_{\rm LSR}\approx213$ \kms\ component is similar to that seen in a number of other LMC sight lines with no apparent correlation to LMC structure \citep{Danforth02}.  We interpret this as either a high-velocity cloud along the line of sight, or a series of neutral LMC halo components.

\citet{Howk02a} found that LMC \OVI\ profiles were systematically shifted $\sim-30$ \kms\ with respect to \FeII\ absorption.  This is consistent with either a halo that corotates with the LMC disk, or a vertical outflow of hot gas from the disk at $v_\bot\sim40$ \kms.  Unfortunately, the $\sim40$ \kms\ expansion velocity of N70 makes SB \OVI\ absorption kinematically indistinguishable from typical halo \OVI\ absorption.  Comparing absolute component velocities to better than $\sim10$ \kms\ is difficult with \FUSE\ data, but we note that the \OVI\ centroid velocity of $242\pm5$ \kms\ is somewhat lower than the 254 \kms\ \FeII\ component assumed to represent the expanding shell.  This is also $\Delta v\sim60$ \kms\ from the assumed systemic velocity.  The comparison sight lines show $v_{\rm OVI}=233\pm15$ \kms.

There is a significant variability in \NOVI\ between different LMC sight lines \citep{Howk02a}.  However, this is the first study to examine closely spaced sight lines through a single ISM structure in the LMC.  Three of the four N70 sight lines show \NOVI\ consistent within their error bars in both the AOD and Voigt fitting methods.  The exception (\sk{67}{250}) shows noticeably stronger absorption.  The uniformity of the three sight lines over $\le87$\arcsec\ ($\le22$ pc at the distance of the LMC) suggests that \OVI\ is distributed evenly throughout N70 on roughly this scale.  However, the \sk{67}{250} sight line is located only 16\arcsec\ ($\sim4$ pc) from the LH\,114-7 sight line and shows an excess \OVI\ column of $\sim0.1$ dex.  This angular scale is similar to that of \Ha\ filaments seen at the edges of N70 and the column density difference may indicate that \OVI\ is distributed nonuniformly on about the same scale.

We now discuss the possible origins of the observed $N_{\rm OVI\sim10^{14}}$~cm$^{-2}$ excess column in N70.  Four possibilities include photoionization, shock heating, turbulent mixing at the interface between the hot interior and the cold shell, or thermal conduction.  Photoionization of O$^{4+}$ requires a harder spectrum (photons of $>114$ eV) than even the hottest O stars can produce\footnote{While O\,VI may be produced locally via photoionization by extremely hot white dwarf stars \citep{DupreeRaymond83}, this is considered unfeasible in large-scale ISM contexts.}.  

\OVI\ is often attributed to shocks, and indeed observed SNR column densities are consistent with the N70 observations \citep[$N_{\rm OVI}=2.5\times10^{14}$~cm$^{-2}$ toward SNR\,0057-7226,][]{Hoopes01}.  However, \citet{Oey00} infer $v_{\rm shock}\sim70$ \kms\ based on optical emission line ratios across the face of N70.  Although this is high for the expansion velocity of a SB, it is far too slow to produce any significant amount of \OVI.  Furthermore, our sight lines intersect the SB shell nearly perpendicular to the presumed direction of expansion, yet we see \OVI\ absorption near the systemic velocity, inconsistent with a shock front fast enough to produce \OVI.

Turbulent mixing layers involve a hot, low-density phase and a warm, high-density phase with a relative motion between them.  Cooler material is entrained and heated by the hot flow in turbulence at all scales and OVI and related ions may be generated.  Given the known expansion of N70 and the convoluted morphology of the optical shell, such relative motions are plausible.  However, current simulations predict very low column densities ($N_{\rm OVI}\la10^{12}$ cm$^{-2}$) per interface over physical scales of a few parsecs \citep{Slavin93,Slavin06}.  While multiple turbulent mixing layers may explain the observed \OVI\ in the halo, there is not enough room for sufficient interfaces within a structure as small as N70.

Thermal conduction models seem to provide the best explanation for the observed \NOVI, but still have difficulty producing sufficient column density.  In this mechanism, a hot, low-density phase ($T\sim10^6$~K) is in contact with a cooler, high-density phase.  Energy from the hot gas propagates into the cooler material as an evaporation front.  When a significant amount of material reaches $T\sim10^5$~K, radiative cooling becomes dominant and the evaporation front stalls.  A condensation front then propagates back into the hot side.  \citet{Borkowski90} model a plane-parallel conduction interface and with \OVI\ present in both evaporation and condensation fronts with a column density $N\rm_{OVI}\approx few\times10^{13}$ cm$^{-2}$.  \citet{Weaver77} model a wind-blown bubble with a conductive interface between the hot, shocked stellar wind and the cool, swept-up ISM and find $N_{\rm OVI}\approx2.3\times10^{13}$ cm$^{-2}$.  \citet{SlavinCox92} calculate $N_{\rm OVI}=1.5-6.8\times10^{13}$ cm$^{-2}$ for old SNRs, depending on a number of parameters.  More recent models of old cooling bubbles come close to $N_{\rm OVI}\sim10^{14}$ cm$^{-2}$ (J. D. Slavin, priv. comm.), but cannot quite explain the observed \OVI\ excess in N70.  It is also unclear if there has been sufficient time in the evolution of N70 for conductive interfaces to generate their maximum potential \NOVI.

With the complexity of the nebular structure of N70 as seen in \Ha, we could invoke multiple interfaces to bring the total \NOVI\ up to the observed values.  However, it seems unlikely that either all four sight lines pierce the same number of strong interfaces (such as conductive interfaces), or that there is enough volume inside N70 for a statistically large number of weak interfaces (such as turbulent mixing layers).  We conclude that a single thermal conduction interface is the most likely source of the observed hot gas, although some combination of these options is also possible.  Observations of other Li-like ions (\NV, \CIV, \SiIV) would be of great help in differentiating heating scenarios \citep{Spitzer96,IndebetouwShull04}, although no instrument is currently available to make the needed observations.  Photoionization will likely also play a role in the production of these lower ions, but many of the critical parameters (spectral type of the photoionizing stars, gas density, etc) are already well-constrained making the modelling relatively benign.

As a further reality check, we estimate the total hydrogen density associated with the \OVI\ implied by our measurements.  Assuming that the \OVI\ is distributed in a shell of thickness $l$, then 
\begin{equation}n_{\rm H}=\frac{N_{\rm OVI}/l}{(O/H)~f_{\rm OVI}}.\end{equation} 
Assuming an \OVI\ fractional ion abundance $f_{\rm OVI}=0.2$ and a typical LMC \HII\ region abundance $(O/H)=2.34\times10^{-4}$ \citep{RussellDopita90}, we derive $n_{\rm H}=0.75/l~\rm cm^{-3}~pc^{-1}$.  For an \OVI\ shell occupying the outer 10\% of the SB radius ($l\sim5$~pc), we get $n\rm_H\sim0.15~cm^{-3}$.  This value is nicely intermediate between the inferred densities of the photoionized shell \citep[$n\rm_{H}\sim10~cm^{-3}$,][]{Oey00} and the hot bubble interior \citep[$n\rm_{H}\sim10^{-3}~cm^{-3}$,][]{OeyGarciaSegura04}, as one would expect from a thermal conduction interface.

In addition to the \OVI\ absorption seen toward the stellar targets, we can search for \OVI\ emission from N70.  The four N70 sight lines are in a crowded stellar field and were observed with the 4\arcsec$\times$20\arcsec\ \FUSE\ MDRS aperture.  The 30\arcsec$\times$30\arcsec LWRS aperture is positioned 208\arcsec\ away (52 pc at the distance of the LMC) from the MDRS aperture at a position angle determined by the \FUSE\ spacecraft roll angle at the time of each observation.  LWRS aperture placements can potentially constrain \OVI\ emission and, in conjunction with column densities measured in absorption, determine the gas density $n_{\rm OVI}$.\footnote{For instance, \citet{Danforth03} detected $F\rm_{OVI}=4.1\times10^{-6}~erg~cm^{-2}~s^{-1}~sr^{-1}$ in a LWRS aperture placed serendipitously on the edge of SMC SNR\,0057$-$7226 in an exposure of only 11 ksec.}  Unfortunately, the LWRS aperture locations for all four N70 pointings fall just outside the bright \Ha\ limb of the SB as shown by the dotted boxes in Figure~1.  Hence, we instead use these observations to place an upper limit on any general diffuse \OVI\ emission in the region near N70.  

None of the individual LWRS pointings show emission in either of the \OVI\ doublet lines at either Galactic or LMC velocities.  Coadding the four LiF1a data sets to increase the S/N, we have a total of 41.7 ks of exposure time.  The coadded data show no evidence for \OVI\ emission and we set a $2\sigma$ upper limit for diffuse \OVI\ emission of $F\rm_{OVI}<1.8\times10^{-7}~erg~cm^{-2}~s^{-1}~sr^{-1}$ or $F\rm_{OVI}<9600~photons~cm^{-2}~s^{-1}~sr^{-1}$.  This upper limit is not as sensitive as the $2\sigma$ limit on diffuse \OVI\ emission in the local bubble of $<500~\rm photons~cm^{-2}~s^{-1}~sr^{-1}$ \citep{Shelton03} but is $\sim20$ times lower than the \OVI\ emission found by \citet{Danforth03} on the limb of SNR\,0057-7226.  A LWRS aperture placed within the ring of N70 could potentially show high-ion line emission and would be a definitive test for SB \OVI.

Finally, we note that the composite N70 and comparison profiles don't match perfectly in Figure~4 in the Galactic component.  The N70 profile shows a slight excess in absorption at $v_{\rm LSR}\approx120$ \kms\ while the profiles at $v\approx0$ \kms\ match quite well.  This excess is nominally in the Galactic absorption component but could plausibly represent hot material escaping from N70 with a projected blueshift of $\sim150$ \kms.  \citet{OeyGarciaSegura04} note that many SBs expand slower than predicted given the mechanical energy input of stellar winds and SN.  They posit that hot gas may be escaping through holes in the photoionized shell thus decreasing the internal pressure.  This process leads to galactic chimneys on larger scales and is thought to be responsible for the general hot phase of the ISM.  A hot component escaping at the observed velocity might be evidence for such an ``\OVI\ leak.''  If confirmed, such a model might increase the viability of shocks contributing significantly to the observed \OVI\ excess from N70.

\section{Conclusions}

We have compared \FUSE\ data for four sight lines through the $\sim105$ pc diameter LMC superbubble N70 with four nearby sight lines sampling diffuse and field regions in the LMC.  This position on the eastern edge of the LMC is isolated from other large-scale ISM structures such as supergiant shells, \HII\ regions, supernova remnants, and other SBs.  Thus, our sight lines are subject to minimal sight line confusion.  We see strong \OVI\ absorption in all eight sight lines organized into two clear absorption components; one at $-50<v_{\rm LSR}<+150$ \kms\ arising in the Galactic halo, and another at $+150<v_{\rm LSR}<+350$ \kms\ from LMC gas.  We measure \OVI\ absorption using both integrated apparent optical depth and profile-fitting methods; both methods yield consistent results.  The Galactic column densities, line widths, and profile shapes are equivalent for the N70 and comparison sight lines and are consistent with typical Galactic \NOVI\ seen in other surveys.  The N70 sight lines show a $\sim0.2$ dex ($\sim60\%$) higher column density at LMC velocities than the comparison sight lines, larger than the scatter in either sample.  This corresponds to $N_{\rm OVI}\approx10^{14}$~cm$^{-2}$.  Furthermore, the N70 \OVI\ profiles show slightly broader absorption than do the comparison profiles.

Our measured \NOVI\ values for both N70 and comparison sight lines are consistent with those found by \citet{Howk02a} in an earlier survey of LMC \OVI\ absorption along twelve sight lines.  While Howk et al. claim no correlation between \NOVI\ and ISM morphology (as measured by either stellar spectral type or \Ha\ or soft X-ray surface brightness), we find that our N70 \NOVI\ is consistent with that measured in four SB sight lines in their sample.  Similarly, our four comparison measurements are entirely consistent with the eight non-SB measurements in the Howk et al. sample.  We show that the SB and non-SB samples show distinct mean \NOVI\ but that the scatter (as expected from halo variation) is equivalent for both samples.  This is strong evidence that SBs (including N70) produce a local reservoir of \OVI\ distinct and distinguishable from the general, overlying halo absorption seen in nearly all LMC sight lines.

We compare \OVI\ absorption to \FeII, a reliable tracer of the warm LMC disk gas.  The strongest \FeII\ absorption traces the systemic velocity of LMC disk gas near N70 at $v_{\rm LSR}=295$ \kms\ and the observed \OVI\ absorption is consistent with an expansion velocity of $\sim60$ \kms.  However, it is also consistent with the general trend of blueshifted halo \OVI\ absorption with respect to \FeII\ seen by \citet{Howk02a}.

We consider four different ionization mechanisms (photoionization, shocks, turbulent mixing layers, and conductive interfaces) and conclude that thermal conduction between the interior hot, X-ray producing gas and the cool, photoionized shell of N70 is the most feasible production mechanism for the observed \OVI.  The inferred shock velocity \citep[$v_{\rm shock}\sim70$ \kms,][]{Oey00} is insufficient to ionize the SB shell to the observed level.  We calculate the total hydrogen density implied by our \NOVI\ measurements and find $n_{\rm H}\sim0.15~\rm cm^{-3}$ for a 5 pc \OVI\ shell, consistent with expected values.

Our LWRS apertures lie just outside the shell of N70 and show no \OVI\ emission to a fairly sensitive level ($F\rm_{OVI}<1.8\times10^{-7}~erg~cm^{-2}~s^{-1}~sr^{-1}$).  Future detections of diffuse \OVI\ emission toward N70 will conclusively demonstrate the existence of \OVI\ within the SB.  In addition, emission-absorption measurements can constrain the density and distribution of hot gas within N70.

\vspace{.3 cm}

It is our pleasure to acknowledge fruitful discussions with Jonathan Slavin, Mike Shull, You-Hua Chu, Ken Sembach, Alex Fullerton, and Ravi Sankrit.  Alex Fullerton provided reduced, archival \FUSE\ data for the comparison sight lines and Chris Smith supplied the interference filters used in creating Figure~1.  POSS-2/UKSTU images in Figure~2 are courtesy of the UK Schmidt Telescope and digitized at the Space Telescope Science Institute under U.S. Government grant NAG W-2166.  This work contains data obtained for the Guest Observer Program by the NASA-CNES-CSA \FUSE\ mission operated by the Johns Hopkins University.  Financial support has been provided by NASA grants NAG5-13689 to University of Colorado and NAG5-13704 and NNG05GE03G to Johns Hopkins University.

\end{document}